\documentclass[preprint,showpacs,preprintnumbers,amsmath,amssymb]{revtex4}
\usepackage{graphicx}
\usepackage{dcolumn}
\usepackage{bm}
\usepackage{hyperref}
\usepackage[usenames]{color} 

\begin{document}

\title{Electronic structure of ruthenium-doped iron chalcogenides}

\author{M. J. Winiarski}
\author{M. Samsel-Czeka\l a}
\affiliation{Institute of Low Temperature and Structure Research, Polish Academy of Sciences,
 	 ul. Ok\'olna 2, 50-422 Wroclaw, Poland}

\author{A. Ciechan}
\affiliation{Institute of Physics, Polish Academy of Sciences,
 	al. Lotnik\'ow 32/46, 02-668 Warsaw, Poland}
 	 
\date{\today}

\begin{abstract}
The structural and electronic properties of hypothetical Ru$_x$Fe$_{1-x}$Se and Ru$_x$Fe$_{1-x}$Te systems have been investigated from first principles within the density functional theory (DFT). Reasonable values of lattice parameters and chalcogen atomic positions in the tetragonal unit cell of iron chalcogenides have been obtained with the use of norm-conserving pseudopotentials. The well known discrepancies between experimental data and DFT-calculated results for structural parameters of iron chalcogenides are related to the semicore atomic states which were frozen in the used here approach. Such an approach yields valid results of the electronic structures of the investigated compounds.
The Ru-based chalcogenides exhibit the same topology of the Fermi surface (FS) as that of FeSe, differing only in subtle FS nesting features.
Our calculations predict that the ground states of RuSe and RuTe are nonmagnetic, whereas those of the solid solutions Ru$_x$Fe$_{1-x}$Se and Ru$_x$Fe$_{1-x}$Te become the single- and double-stripe antiferromagnetic, respectively. However, the calculated stabilization energy values are comparable for each system. The phase transitions between these magnetic arrangements may be induced by slight changes of the chalcogen atom positions and the lattice parameters {\it a} in the unit cell of iron selenides and tellurides. Since the superconductivity in iron chalcogenides is believed to be mediated by the spin fluctuations in single-stripe magnetic phase, the Ru$_x$Fe$_{1-x}$Se and Ru$_x$Fe$_{1-x}$Te systems are good candidates for new superconducting iron-based materials.

\vskip 0.2cm
\noindent 

\end{abstract}

\maketitle

\section{\label{sec1}Introduction}

An interplay between magnetism and superconductivity (SC) in iron chalcogenides still draws wide interest.\cite{Dagotto} In FeSe a rapid increase of the superconducting critical temperature ($T_c$) from 8~K in equilibrium conditions \cite{Hsu} up to 37~K under hydrostatic pressure \cite{Mizuguchi-FeSe,Margadonna,Medvedev,Okabe} has been reported. The  solid solutions FeSe$_{1-x}$Te$_x$ are superconducting for $x<$0.8 with the maximum $T_c=$15~K observed for $x=$0.5,\cite{Yeh,Fang,Gawryluk} whereas in FeTe$_{1-x}$S$_x$ the value of $T_c$ reaches 10 K for $x=$0.2.\cite{Mizuguchi_FeTeS,Hu,Zajdel,Awana,Dong}

The SC phenomenon in iron chalcogenides is somewhat connected with the changes of the tetrahedral coordination of Fe atoms, \cite{Medvedev,Okabe,Han} thus the $T_c$ of particular materials can be modified by non-hydrostatic pressure in lattice mismatched epitaxial films. The tensile strain suppresses superconductivity of FeSe on MgO and SrTiO$_3$, \cite{Nie} while the compressive biaxial ($ab$-plane) or uniaxial ($c$-axis) strain on Fe(Se,Te) causes the increase of the $T_c$'s.\cite{Huang,Bellingeri,Si} Interestingly, in opposition to FeSe the SC in FeTe emerges in tensile-strained thin films.\cite{Han}   

The electronic structure of FeSe-based superconductors has been extensively investigated in both experimental \cite{Chen,Nakayama,Tamai,Miao,Perez} and theoretical studies. \cite{Subedi,Ding,Singh,Chadov,APPa,EPL,JAlloyCompd,Intermet_FeSe,Intermet,JPhysCondMatt} It is believed that the multi-gap SC in 11-type compounds originates from the interband interactions between the holelike $\beta$ and electronlike $\delta$ Fermi surface (FS) sheets. In particular, SC can be mediated by antiferromagnetic (AFM) spin fluctuations,\cite{NMR,Qiu} which are driven by the imperfect nesting with the $\mathbf{q}\approx(0.5,0.5)\times(2\pi /a)$ vector, spanning the above FS sheets in the iron chalcogenides.\cite{Subedi,Singh,APPa,EPL,JAlloyCompd,Intermet}
Furthermore, such fluctuations are related to the single-stripe AFM order, while compounds with the double-stripe AFM order do not exhibit SC.\cite{Liu} The magnetic ordering in iron chalcogenides is closely connected with the chalcogen atom position in the unit cell.\cite{Moon,JPhysCondMatt}

Problems with reconciling density functional theory (DFT) calculations with experiment in ferropnictides have been already extensively discussed.\cite{Mazin} However, the issue of a structural optimization for iron-based layered compounds is generally solved by the use of experimental lattice parameters, whereas some authors tested a performance of the van der Waals interaction corrections. \cite{Ricci,Ye}

In this work we show that the use of norm-conserving pseudopotentials in the standard LDA approach may lead to reasonable results of structural properties for iron chalcogenides. Since no experimental investigations of the studied here ruthenium chalcogenides have been reported so far, the examination of quality of our DFT-based predictions is a crucial issue.

The electronic structure modifications following from the substitution of Fe with Ru atoms in the solid solutions Ru$_x$Fe$_{1-x}$Se and Ru$_x$Fe$_{1-x}$Te  are analyzed with special attention paid to possible implications for SC phenomenon in these hypothetical materials. Namely, the single-stripe antiferromagnetic ground state and the topology of the FS with the corresponding imperfectly nested area between the $\beta$ and $\delta$ sheets can be responsible for a spin-fluctuation mediated superconducting pairing in such systems.
Since in iron pnictides the doping with the Ru atoms raise the SC,\cite{Sharma,Brouet,Dhaka,Xu,Ma} this study should encourage further experimental investigations of Ru-doped iron chalcogenides. 

\section{\label{sec2}Computational methods}
Band structure calculations for iron and ruthenium chalcogenides have been carried out in the framework of DFT. A full optimization of the free $z_{Se/Te}$ atomic positions and geometry of the unit cell (u.c.)  was performed with the Abinit package,\cite{Abinit} using the norm-conserving pseudopotentials, generated with APE software.\cite{APE} The local density approximation (LDA) \cite{JP} of the exchange-correlation potential was employed. The $3d4s4p$ states for Fe and Se pseudoatoms as well as the $4d5s5p$ states for the Ru and Te pseudoatoms were selected as a valence-band basis.
Calculations of an equilibrium volume of u.c. for parent compounds (FeSe, FeTe, RuSe, RuTe) were performed in the tetragonal PbO-type phase in nonmagnetic mode. Then, Ru$_x$Fe$_{1-x}$Se and Ru$_x$Fe$_{1-x}$Te systems for $x = 0.25, 0.5, 0.75$ were simulated with the tetragonal supercells (8-atoms in $\sqrt{2}a \times \sqrt{2}a \times c$ multiplication of the primitive u.c.) of the PbO-type.
In spin-polarised calculations (LSDA) the ($\sqrt{2}a \times \sqrt{2}a \times c$) and ($2a \times a \times c$) supercells were employed for single- (AFM1) and double-stripe (AFM2) antiferromagnetic orders, respectively. A further relaxation of atomic positions (anion height $z_{Se/Te}$) and the shape of u.c. of those systems were performed starting from volumes of u.c. obtained in the former nonmagnetic calculations. Such a relaxation leads to the orthorhombic and monoclinic distortions of u.c., for AFM1 and AFM2 ordered systems. The calculated values of the magnetic stabilization energy are related to the volume of u.c. in the non-magnetic phase.

Based on the optimized structural properties of the parent FeSe, FeTe, RuSe, and RuTe compounds, the full potential local-orbital (FPLO) band structure code \cite{FPLO} was used to compute the densities of states (DOSs) and Fermi surfaces in the nonmagnetic phase (the tetragonal u.c. of PbO-type). Since the FS nesting features  of the 11-type systems are tiny, very dense {\bf k}-point meshes in the BZ had to be used, {\it i.e.} 64$\times$64$\times$64 and 256$\times$256$\times$256 for the self-consistent field (SCF) cycle and FS maps, respectively.

Finally, a nesting function was determined numerically by the formula: 

\begin{equation}
f_{nest}(\mathbf{q})=\Sigma_{\mathbf{k},n,n'}
\frac{[1-F_{n}^{\beta}({\mathbf{k}})]F_{n'}^{\delta}(\mathbf{k+q})}
{|E_{n}^{\beta}({\mathbf{k}})-E_{n'}^{\delta}(\mathbf{k+q})|},
\end{equation}

where $F_{n}^{\beta}$ and $F_{n'}^{\delta}$ are the Fermi-Dirac functions of states $n$ and $n'$ in bands $\beta$ and $\delta$, ($F=$ 0 or 1 for holes or electrons), respectively. $E_{n}^{\beta}$ and $E_{n'}^{\delta}$ are energy eigenvalues of these bands. The studied $f_{nest}({\mathbf{q||Q}})$, were $\mathbf{Q}=(0.5,0.5)\times(2\pi /a)$ is the ideal nesting vector, represents a frequency of an occurrence of a given vector $\mathbf{q}\sim(\pi,\pi)$ (having its length close or equal to that of $\mathbf{Q}$) in the {\bf k}-space, spanning the FS sheets originating from the $\beta$ and $\delta$ bands. It is worth noting that the calculated $f_{nest}({\mathbf{q||Q}})$ is not exactly equivalent to the Lindhard susceptibility. Furthermore, such a simple form of the $f_{nest}$ is insufficient for an estimation of the effective pairing interaction in multiorbital systems. For an extensive study of the spin-fluctuation mediated pairing in Fe-based compounds see e.g. \cite{Graser}.

\section{Results and discussion}

The calculated structural parameters $a$, $c/a$, and $z_{Se/Te}$ of the tetragonal PbO-type u.c. of iron and ruthenium chalcogenides, compared to the literature data, are gathered in Table \ref{table1}. Generally, the DFT-derived values of the lattice parameter {\it a} are underestimated whereas the values of {\it c/a} ratio are strongly overestimated within both GGA and LDA approaches.
This issue is related to an anisotropic crystal structure of the studied here systems, in which metallic layers formed by iron and chalcogen atoms, being perpendicular to the elongated {\it c} axis, are connected to one another with bonds that can be effectively described by more sophisticated methods including the van der Waals interaction corrections.\cite{Ricci}

Interestingly, the careful comparison between the LDA results reported in our former studies,\cite{EPL,JAlloyCompd,Intermet} obtained with PAW pseudoatoms containing the extended valence sets, and presented here structural data, calculated with the norm-conserving pseudopotentials, has shown that the lack of semicore states (e.g. Fe $3s3p$ electrons) in the later approach may also lead to reasonable agreement with the experimental values of lattice parameters of iron chalcogenides.
This effect suggests that the problem of the overestimated distance between Fe-Se or Fe-Te layers is related to an incorrect hybridization of semicore states, whereas the lattice parameters {\it a} in this family of compounds are less affected by this phenomenon.
Therefore, the pseudopotential approach may be a useful tool for studies of systems with such anisotropic crystal structures.

The LDA potential seems to be universal for investigations of both selenides and tellurides, while the GGA results \cite{Ricci} for FeSe are clearly insufficient due to the significant overestimation of the {\it c/a} ratio. Furthermore, the magnetic phase diagrams of Fe-based superconductors should be better described by the LDA than the GGA approach.\cite{Mazin} 
Although the spin-polarized calculations lead to somewhat better structural results for FeSe, in the case of FeTe the obtained $c/a$ and $z_{Se/Te}$ are relatively too high when compared to the available  experimental data. Thus, the equilibrium structures of the studied systems may be better estimated by the non-magnetic calculations.

\begin{figure}
\includegraphics[scale=1.0]{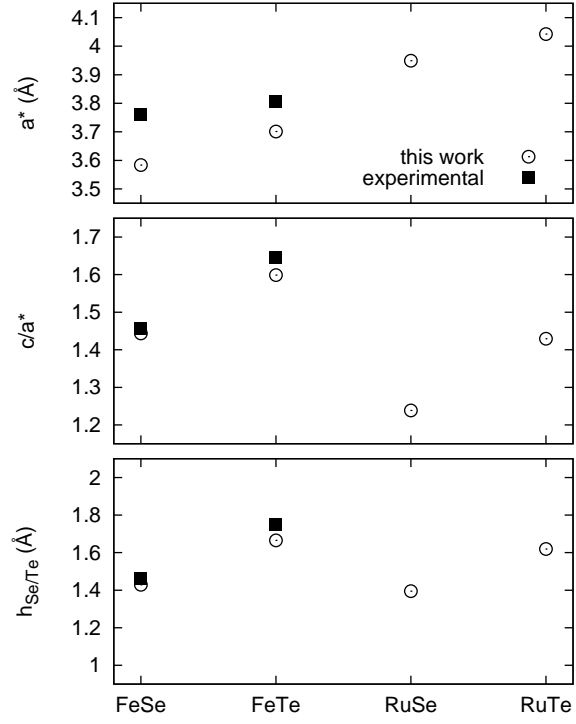}
\caption{Calculated structural parameters $a$, $c/a$, $h_{Se/Te}$ for iron and ruthenium chalcogenides for the PbO-type unit cell. Available experimental data for low-temperature orthorhombic and monoclinic structures are taken from Refs. \cite{Louca,Martinelli}. Note, that for the experimental structures $a^*$ denotes $(a+b)/2$ while $c/a^*$ denotes $2c/(a+b)$.}
\label{Fig1}
\end{figure}

Since no experimental investigations of ruthenium chalcogenides have been reported until now, the predicted here structural parameters {\it a}, {\it c/a}, and the chalcogen atom height $h_{Se/Te}$ of Ru- and Fe-based compounds, presented in Table \ref{table1} and Fig. \ref{Fig1}, require some further discussion. The lengths of the lattice parameters {\it a} in the tetragonal PbO-type structure are somehow related to the size of particular transition metal atom, thus a difference of about~0.35~{\AA} between calculated values of {\it a} for Fe- and Ru-based systems as well as the analogous change of {\it a} by about 0.01~{\AA} between Se- and Te-based compounds have been revealed, as seen in Fig. \ref{Fig1}.
Furthermore, the {\it c/a} ratio and $h_{Se/Te}$, also exhibit similar chemical trends, though the difference in their values between Fe- and Ru-based compounds is negative. It is worth noting that despite the calculated values of the lattice parameter {\it a} for ruthenium chalcogenides are relatively high, the positions of the chalcogen atoms in u.c. of these systems, are close to those of iron chalcogenides. Namely, the values of $h_{Te}$ in FeTe and RuTe are almost the same.

\begin{figure}
\includegraphics[scale=1.0]{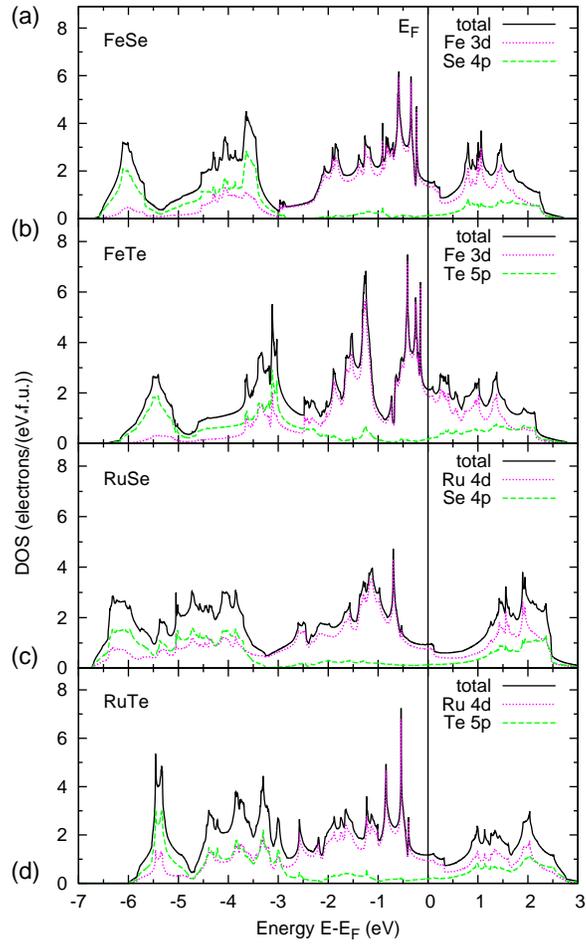}
\caption{The total and orbital projected electronic DOS (LDA) for FeSe (a), FeTe (b), RuSe (c), and RuTe (d).}
\label{Fig2}
\end{figure}

Next, the calculated DOS of parent iron and ruthenium chalcogenides are presented in Fig. \ref{Fig2}. The overall shapes of the total DOS for FeSe and FeTe are similar to those reported earlier,\cite{Subedi,Ding,Singh,APPa,EPL,JPhysCondMatt,Intermet} being dominated by the Fe $3d$ and Se/Te $4p/5p$ states. For FeSe the obtained here value of DOS at the Fermi level, $N(E_F)$ = 1.57 electrons/eV/f.u., is close to former results,\cite{APPa, EPL} although it remains significantly higher than the value of 0.95 electrons/eV/f.u. reported for the experimental lattice parameters.\cite{Subedi} In the case of FeTe, calculated $N(E_F)$ = 2.14 electrons/eV/f.u. is lower than the value of 2.59 electrons/eV/f.u. \cite{Intermet} for optimized structure, although being still higher than 1.83 electrons/eV/f.u. for the experimental one.\cite{Subedi}
Ruthenium compounds exhibit a similar overall shape of the total DOS to that of FeSe, however, the predicted low $N(E_F)$ of 0.96 and 1.17 electrons/eV/f.u. for RuSe and RuTe, respectively, suggest a relatively weaker metallic character when compared to that of iron chalcogenides. Note that the substitution of Fe with Ru atoms leads to diminishing of $N(E_F)$ also in other Fe-based intermetallics, e.g. superconducting Lu$_2$Fe$_3$Si$_5$.\cite{Lu2Fe3Si5}

\begin{figure}
\includegraphics[scale=1.0,angle=-90]{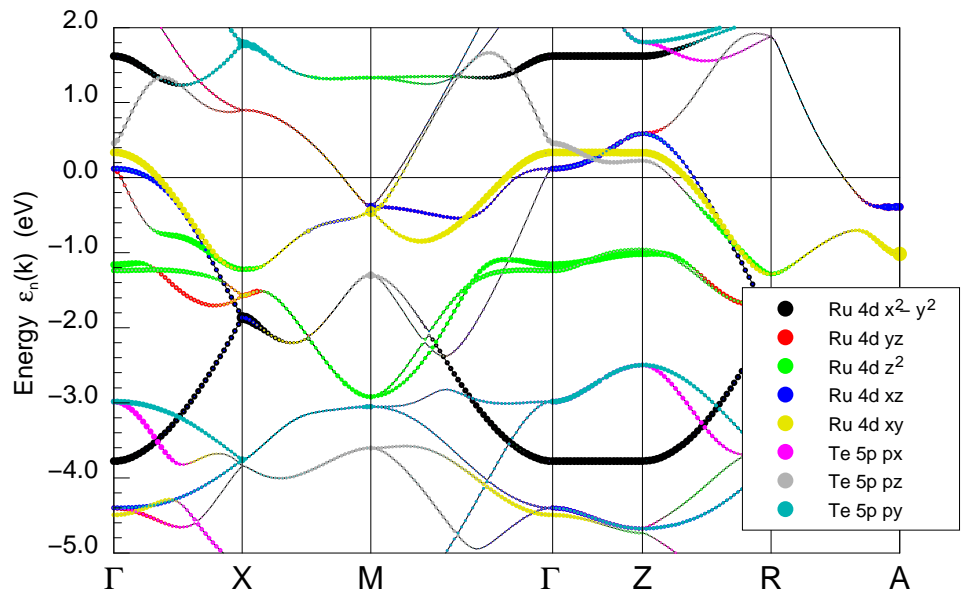}
\caption{Computed (LDA) weighted bands for RuTe. The selected here predominant Ru $4d$ and Te $5p$ orbital characters are marked by circles of different colors. The circle sizes are proportional to given band weights of the orbitals.}
\label{Fig3}
\end{figure}

Unconventional SC phenomenon in iron compounds is suspected to be related to the pairing between the Fe $3d$ orbitals of specific symmetry types,\cite{Graser} being present also in materials exhibiting low values of $T_c$, e.g. YFe$_2$Ge$_2$,\cite{YFe2Ge2} though being less distinct in compounds with a more isotropic crystal structure, e.g. Lu$_2$Fe$_3$Si$_5$.\cite{Lu2Fe3Si5} This specific band structure character is also present in ruthenium chalcogenides, as illustrated by weighted bandplots for RuTe in Fig. \ref{Fig3}.
Similarly to the former experimental and theoretical results for FeSe,\cite{Maletz,Eschrig} the first two holelike FS elements centered at the $\Gamma$ point are formed by a mixture of the $d_{xz}$ and $d_{yz}$ orbitals, whereas the third is complitely dominated by the $d_{xy}$ orbitals. The two electronlike FS elements around the $M$ point exhibit analogous orbital characters. Thus, the substitution of Fe atoms with Ru atoms should not change the orbital symmetry of the Fermi surface sheets of iron chalcogenides.

\begin{figure}
\includegraphics[scale=0.9]{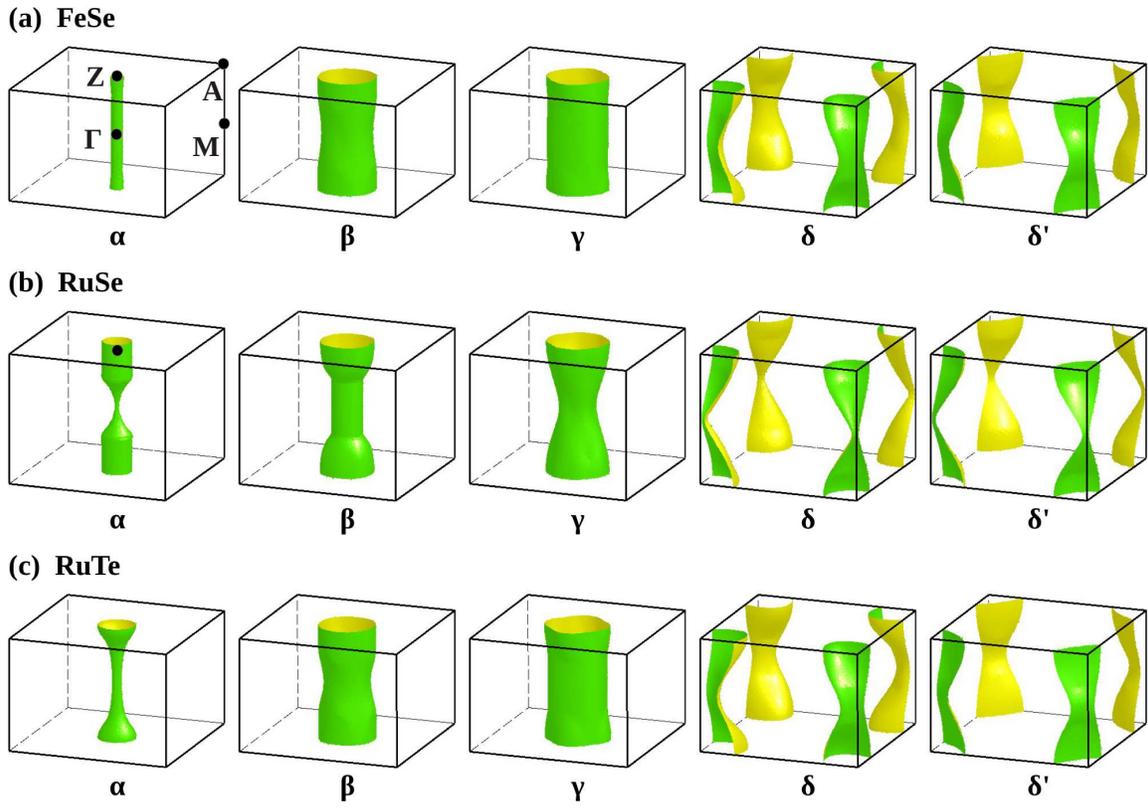}
\caption{The calculated (LDA) holelike ($\alpha$, $\beta$, $\gamma$) and electronlike ($\delta$, $\delta'$) Fermi surface sheets for a) FeSe, b) RuSe, and c) RuTe.}
\label{Fig4}
\end{figure}

\begin{figure}
\includegraphics[scale=1.0]{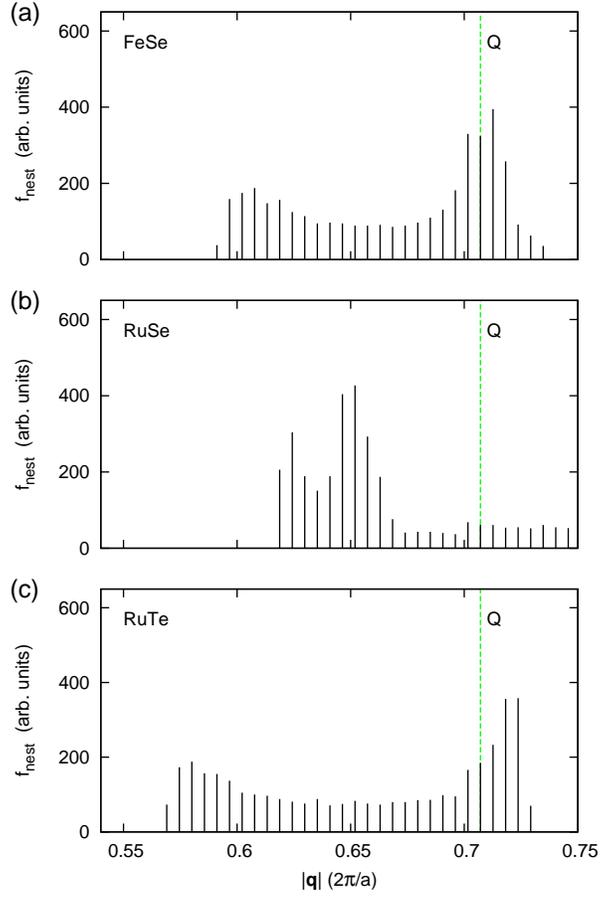}
\caption{Histograms representing the nesting function, $f_{nest}$ vs. lengths of possible vectors ${\mathbf{q||Q}}$ spanning $\beta$ and $\delta$ FS sheets of FeSe (a) RuSe (b), and RuTe (c).}
\label{Fig5}
\end{figure}

Interestingly, the calculated Fermi surface (FS) sheets of RuSe and RuTe, depicted in Fig. \ref{Fig4}, exhibit the same topology as that of FeSe. Furthermore, one can find close similarities between the shape of FS sheets of RuSe and uniaxially compressive strained FeSe \cite{EPL}, whereas the FS of RuTe resembles that of FeSe under high (8-9 GPa) hydrostatic pressure.\cite{APPa} These findings have been examined in detail by the calculations of a nesting function, and focusing on the intensity of vectors $\mathbf{q}\approx(0.5,0.5)\times(2\pi /a)$, as presented in Fig. \ref{Fig5}.
Such a nested area of FS in RuSe is negligible, thus the spin-fluctuation mediated SC in the Ru$_x$Fe$_{1-x}$Se systems with high Ru content $x$ is rather impossible.
An analogous effect has been considered for the tensile strained FeSe in Ref. \cite{EPL} and, indeed, the tensile-strained thin films of FeSe do not exhibit SC.\cite{Nie} However, the same analysis of the FS nesting for RuTe leads to opposite conclusions, SC can be raised in the Ru$_x$Fe$_{1-x}$Te compounds similarly to results reported for tensile-straned FeTe.\cite{Intermet} Namely, the intensity of $f_{nest}$ for the ideal vector $\mathbf{Q}$ is diminished, though the overall shape of $f_{nest}$ is similar to that of FeSe.

\begin{figure}
\includegraphics[scale=1.0]{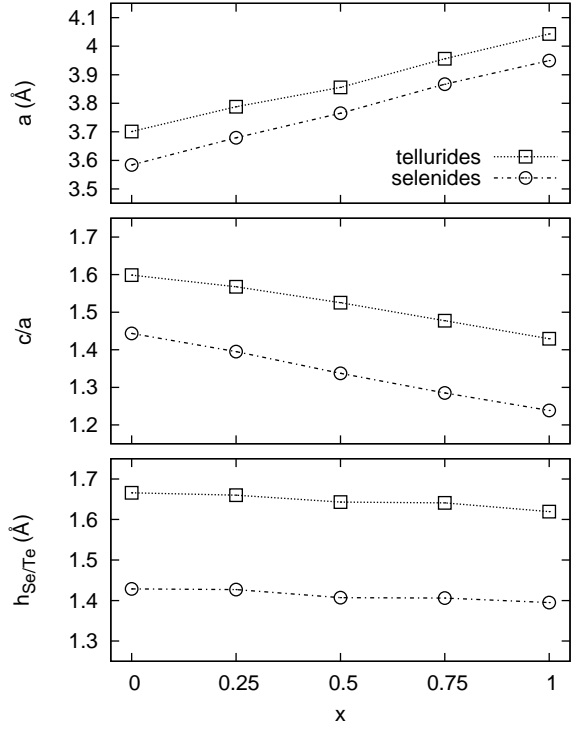}
\caption{Calculated structural parameters {\it a}, {\it c/a}, and $h_{Se/Te}$ for Ru$_x$Fe$_{1-x}$Se and Ru$_x$Fe$_{1-x}$Te systems for the nonmagnetic PbO-type unit cell.}
\label{Fig6}
\end{figure}

The predicted structural parameters and electronic properties of RuSe and RuTe suggest that the Ru-doped iron chalcogenides are possible candidates for new superconductors. The structural data for the Ru$_x$Fe$_{1-x}$Se and Ru$_x$Fe$_{1-x}$Te systems are presented in Fig. \ref{Fig6}. In both materials, the substitution of Fe with Ru atoms leads to a linear increase of the lattice parameter {\it a} and simultaneous decrease of the {\it c/a} ratio.
These specific modifications of a crystal structure, followed from the chemical pressure, cannot be introduced by any kind of strain investigated in former studies for superconducting iron chalcogenides.\cite{APPa, EPL, JAlloyCompd, JPhysCondMatt} The positions of chalcogen atoms in u.c. of Ru$_x$Fe$_{1-x}$Se and Ru$_x$Fe$_{1-x}$Te systems are almost constant with the increase of the Ru content.
Despite the fact that in Fe-based compounds the distance between Fe and chalcogen atoms is somehow connected with a particular magnetic ordering,\cite{Moon,JPhysCondMatt} in Ru-based systems this problem is complicated by the significant increase of the lattice parameter {\it a}, thus some further predictions of a magnetic order in the Ru$_x$Fe$_{1-x}$Se and Ru$_x$Fe$_{1-x}$Te compounds are required.

The enhanced electron-electron correlations between the Fe 3d states in iron (oxy)pnictides and chalcogenides are driven by Hund’s rule coupling rather than by the on-site Hubbard repulsion.\cite{Johannes,Aichhorn} In FeSe the electron doping restores Fermi-liquid properties whereas hole doping enhances bad-metallic properties.\cite{Liebsch} Thus, the substitution of Fe atoms with Ru atoms in iron chalcogenides should lead to the decrease of Hund's coupling and the diminishing of magnetic interactions.

\begin{figure}
\includegraphics[scale=1.0]{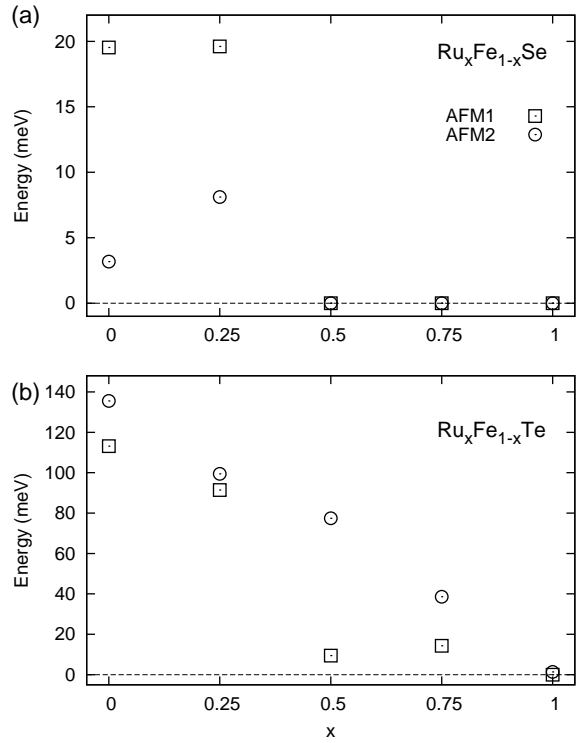}
\caption{Magnetic stabilization energy (vs nonmagnetic) for single- (AFM1) and double-stripe (AFM2) antiferromagnetic orders for Ru$_x$Fe$_{1-x}$Se (a) and Ru$_x$Fe$_{1-x}$Te (b) systems.}
\label{Fig7}
\end{figure}

The calculation results of magnetic stabilization energy for the above systems indicate that the FeSe exhibits a ground state of the single-stripe AFM order (AFM1), as can be seen in Fig. \ref{Fig7} a). Since RuSe is nonmagnetic, the magnetism of Ru$_x$Fe$_{1-x}$Se solid solution diminishes with the increase of the Ru content. An analogous effect observed in iron-pnictides induces the SC for $x \approx$ 0.5,\cite{Sharma,Brouet,Dhaka,Xu,Ma} however, one can expect that for the relatively small u.c. of iron chalcogenides the Ru atoms introduce rapider modifications of the electronic structure. The presented here predictions point out that in Ru$_x$Fe$_{1-x}$Se the values of magnetic stabilization energy are relatively low and the AFM1 magnetic structure is unstable even in the pure FeSe.
These effects may be also explained by significant decrease of the FS nesting intensity in Ru$_x$Fe$_{1-x}$Se systems. All these findings support the idea, that Ru-doped FeSe may be a good candidate material for a new superconductor.

The calculated values of magnetic stabilization energy for tellurides (see Fig. \ref{Fig7} b)) are high compared to those of Ru$_x$Fe$_{1-x}$Se systems, however, the total energy difference between AFM2 and AFM1 phase for FeTe is also relatively low. Despite the fact that the electronic structure of RuTe is very similar to that of FeSe/FeTe superconductors, in the Ru$_x$Fe$_{1-x}$Te materials the relatively high values of lattice parameter {\it a} promote a stable AFM2 order, which is not optimal for an occurrence of spin-fluctuation mediated SC. Therefore, the Ru-doped FeTe appears to be nonsuperconducting.
On the other hand, one can consider some strain-induced changes of the magnetic ordering in Ru$_x$Fe$_{1-x}$Te for Ru contents of about 0.25, analogous to the effects reported for pure FeTe,\cite{JPhysCondMatt} that may raise SC. However, such investigations are beyond the scope of this work. Furthermore, the calculated ground state for $x$ = 0.25 may be also affected by the relaxation of structural parameters within the {\it ab initio} approach. This issue cannot be discussed yet since no experimental data have been available up to now.

\section{\label{sec3}Conclusions}

Structural, magnetic, and electronic properties of novel Ru$_x$Fe$_{x-1}$Se and Ru$_x$Fe$_{x-1}$Te systems have been studied by {\it ab initio} calculations.
The reasonable values of lattice parameters and chalcogen atomic positions in unit cells of superconducting iron chalcogenides can be effectively obtained with the LDA exchange-correlation functional in the pseudopotential approach. The Ru-based systems exhibit the same topology of the Fermi surface as that of FeSe with close similarities to those of strained FeSe and FeTe.
The calculated density of states at the Fermi level for Ru-based compounds is relatively low when compared to those of iron chalcogenides. Since the single-stripe antiferromagnetic order promoting superconductivity is favorable in Ru$_x$Fe$_{x-1}$Se and also possible in Ru$_x$Fe$_{x-1}$Te systems, the Ru-doped FeSe and FeTe compounds seem to be good candidates for new superconducting iron-based materials.

\section*{Acknowledgments}
This work was supported by the National Science Center of Poland under grants No. 2012/05/B/ST3/03095 and No. 2013/08/M/ST3/00927. Calculations were partially performed on ICM supercomputers of Warsaw University (Grant No. G46-13) and in Wroclaw Center for Networking and Supercomputing (Project No. 158).


\begin{table}
\caption{Calculated lattice parameters {\it a} and {\it b}, {\it 2c/(a+b)} ratio, and free atomic positions, $z_{Se/Te}$, in fully optimized u.c. of iron and ruthenium chalcogenides for the nonmagnetic PbO-type (tetragonal) and single- (AFM1, orthorhombic) and double-stripe (AFM2, monoclinic) phases for FeSe and FeTe, respectively. The $\Delta$ values are derived from the available experimental data for FeSe and FeTe.}
\label{table1}
\begin{tabular}{lrrrrrrrr} \\ \hline
reference & {\it a} (\AA) & $\Delta$ (\%) & {\it b} (\AA) & $\Delta$ (\%) &  {\it 2c/(a+b)} & $\Delta$ (\%) &  $z_{Se/Te}$ & $\Delta$ (\%) \\ \hline
FeSe: & & & & & & & & \\
this work LDA & 3.584 & -4.82 & 3.584 & -4.53 & 1.444 & -0.94 & 0.276 & 3.47 \\
this work LSDA (AFM1) & 3.602 & -4.34 & 3.550 & -5.43 & 1.464 & -0.75 & 0.280 & 4.85 \\
ref. \cite {EPL} LDA (PAW) & 3.592 & -4.60 & 3.592 & -4.32 & 1.499 & 2.34 & 0.257 & -3.69 \\
ref. \cite{Ricci} GGA & 3.680 & - 2.27 & 3.680 & -1.97 & 1.701 & 16.15 & 0.222 & -16.81 \\
ref. \cite{Ricci} DFT-D2 & 3.640 & -3.33 & 3.640 & -3.04 & 1.489 & 1.67 & 0.258 & -3.21 \\
ref. \cite{Louca} experimental & 3.765 & 0.00 & 3.754 & 0.00 & 1.457 & 0.00 & 0.267 & 0.00 \\ \hline
FeTe: & & & & & & \\
this work LDA & 3.701 & -3.40 & 3.701 & -1.42 & 1.599 & -2.84 & 0.282 & 0.83 \\
this work LSDA (AFM2) & 3.617 & -5.59 & 3.617 & -3.66 & 1.712 & 4.02 & 0.287 & 2.66 \\
ref. \cite {Intermet} LDA (PAW) & 3.748 & -2.18 & 3.748 & -0.17 & 1.783 & 8.36 & 0.266 & -4.73 \\
ref. \cite{Ricci} GGA & 3.810 & -0.55 & 3.810 & 1.49 & 1.711 & 3.99 & 0.244 & -12.64 \\
ref. \cite{Ricci} DFT-D2 & 3.770 & -1.60 & 3.770 & 0.42 & 1.600 & -2.76 & 0.264 & -5.55 \\
ref. \cite{Martinelli} experimental & 3.831 & 0.00 & 3.783 & 0.00 & 1.645 & 0.00 & 0.279 & 0.00 \\ \hline
RuSe: & & & & & & & & \\
this work LDA & 3.949 & & & & 1.239 & & 0.285 & \\ \hline 
RuTe & & & & & & & & \\
this work LDA & 4.042 & & & & 1.429 & & 0.280 & \\ \hline
\end{tabular}
\end{table}

\end{document}